\documentclass[aps,prd,superscriptaddress,twoside,twocolumn,nofootinbib,showpacs,floatfix]{revtex4-1}
\usepackage{amsmath,amssymb}
\usepackage{graphicx,bm,braket}
\usepackage{subfigure}
\pdfpagewidth=\paperwidth
\pdfpageheight=\paperheight
\allowdisplaybreaks
\usepackage{epstopdf}
\newcommand*{\slashed}[1]{{#1\!\!\!/}}
\newcommand*{\hc}{\text{H.\,c.}}

\begin{document}

\title{\boldmath Combined analysis of the $\gamma n \to K^0\Sigma^0$ and $\gamma n \to K^+\Sigma^-$ reactions}

\author{Neng-Chang Wei}
\affiliation{School of Nuclear Science and Technology, University of Chinese Academy of Sciences, Beijing 101408, China}

\author{Ai-Chao Wang}
\affiliation{School of Nuclear Science and Technology, University of Chinese Academy of Sciences, Beijing 101408, China}

\author{Fei Huang}
\email{Corresponding author. Email: huangfei@ucas.ac.cn}
\affiliation{School of Nuclear Science and Technology, University of Chinese Academy of Sciences, Beijing 101408, China}

\date{\today}

\begin{abstract}
The recently released data on differential cross sections for $\gamma n \to K^0\Sigma^0$ from the A2 and BGOOD Collaborations are used to examine the theoretical model constructed in our previous work [Phys. Rev. D \textbf{105}, 094017 (2022)] for $\gamma n \to K^+\Sigma^-$, and it is found that the model predictions are able to qualitatively reproduce the A2 data but fail to describe the BGOOD data. Then, a combined analysis of the $\gamma n \to K^0\Sigma^0$ and $\gamma n \to K^+\Sigma^-$ reactions is performed to revise the theoretical model. Due to the inconsistency problem, the A2 and BGOOD data are included in fits separately. In the case of including the A2 data, both the data for $\gamma n \to K^0\Sigma^0$ and $\gamma n \to K^+\Sigma^-$ can be fairly well described, and the contributions from the $N(1710)1/2^+$, $N(1880)1/2^+$, $N(1900)3/2^+$, and $\Delta(1920)3/2^+$ resonances are found to dominate the reactions in the lower energy region. While in the case of including the BGOOD data, although most of the data for the $\gamma n \to K^+ \Sigma^-$ reaction can be described with the exception of some noticeable discrepancies on beam asymmetries at lower energies, the BGOOD data for $\gamma n \to K^0\Sigma^0$ can be only qualitatively described, and the contributions from the $N(1710)1/2^+$, $N(1900)3/2^+$, and $\Delta(1910)1/2^+$ resonances are found to dominate the reactions in the lower energy region. In both cases, the $t$-channel $K^\ast(892)$ exchange is found to play a crucial role at forward angles in the higher energy region. Further precise measurements of data for $\gamma n \to K^0\Sigma^0$ are called on to disentangle the discrepancies between the data sets from the A2 and BGOOD Collaborations.
\end{abstract}



\maketitle

\section{Introduction}   \label{Sec:intro}

Studies of the spectrum and structure of the excited states of the nucleon and $\Delta$ ($N^\ast$s and $\Delta^\ast$s) can provide us with essential information about the dynamics of the strong interaction in the nonperturbative regime of quantum chromodynamics (QCD). Extracting information on resonances from the data of the $\pi N$ scattering experiments and single-pion photoproduction experiments represents the most important part of the preceding exploration of the $N^\ast$s and $\Delta^\ast$s. However, both the QCD-inspired phenomenological models \cite{Isgur:1977ef, Koniuk:1979vy} and lattice QCD calculations \cite{Edwards:2013, Lang:2017, Kiratidis:2017, Andersen:2018} have predicted many more $N^\ast$s and $\Delta^\ast$s than those have been unraveled in experiments as listed in the Review of Particle Physics (RPP) \cite{Zyla:2020zbs}, which is known as the ``missing resonance problem'' in baryon spectroscopy.

Since the $\pi N$ channels are sensitive to resonances with sizable substantial decays to the $\pi N$ final states, those ``missing resonances'' may have escaped from observation in the $\pi N$ scattering experiments and/or single-pion photoproduction experiments due to their small couplings to the $\pi N$ final states. In this regards, the hadrons photo- and electroproduction processes can offer opportunity to study resonances with sizable decays not only to $\pi N$ final states but also to other final states such as $\eta N$, $\omega N$, $\rho N$, $K \Lambda$, $K \Sigma$, {\it et al.}. Consequently, studying the spectrum and structure of $N^\ast$s and $\Delta^\ast$s through probes of real and virtual photons has been a key objective of the hadron physics community in the past 20 years \cite{Mokeev:2022afv, Ireland:2019uwn, Laget:2019tou, Aznauryan:2011qj}. Actually, the accumulated high-quality data of various exclusive meson photo- and electroproduction experiments have helped to identify several long-awaited new resonances known previously as the ``missing resonances'' \cite{Mokeev:2022afv}.

Due to the broad and overlapping nature of light-flavor resonances, the identification of individual resonance contributions requires analyzing the data of different exclusive meson photo- and electroproduction reactions that have selective sensitivity to different resonances. Stimulated by the recently released data on beam asymmetries $\Sigma$ \cite{CLAS-beam} and  beam-target asymmetries $E$ \cite{Zachariou:2020kkb} for the $\gamma n \to K^+\Sigma^-$ reaction, in our previous work \cite{Wei:2022nqp} we have performed a timely theoretical analysis of all the available data \cite{Kohri:2006yx, AnefalosPereira:2009zw, CLAS-beam, Zachariou:2020kkb} for the $\gamma n \to K^+\Sigma^-$ reaction based on an effective Lagrangian approach in the Bonn approximation. There, all the available data for the $\gamma n \to K^+\Sigma^-$ reaction were well described and explained by the model results. The reaction mechanisms were analyzed, and the parameters of the relevant resonances were extracted and compared to those quoted by RPP \cite{Zyla:2020zbs}.

Meanwhile, in Ref.~\cite{Byd2021} the same data for the $\gamma n \to K^+ \Sigma^-$ reaction were also analyzed within an isobar model with two sets of fit results, i.e., fit M and fit L, which were obtained by using, respectively, the standard MINUIT and the so-called Least Absolute Shrinkage and Selection Operator method for fitting the data. The data for the $\gamma n \to K^+ \Sigma^-$ reaction can be also well described with the two sets of fit results in Ref.~\cite{Byd2021}, but different conclusions about the reaction mechanisms, in particular about the resonance contents and major resonance contributions, were drew.

The fact that the two independent analyses of Ref.~\cite{Wei:2022nqp} and Ref.~\cite{Byd2021} have led to quite different results indicates that the currently available data for $\gamma n \to K^+ \Sigma^-$ are insufficient to uniquely determine the reaction amplitudes in theoretical models. In fact, even a \textit{mathematically complete} experiment for a pseudoscalar-meson photoproduction process would require data on at least eight carefully chosen independent observables to determine the corresponding reaction amplitudes at each center-of-mass energy and meson polar angle \cite{Chiang:1996em,Nakayama:2019cys,Nakayama:2018yzw}. In this sense, further experimental constraints are needed in fully determining the reaction mechanisms of $\gamma n \to K^+ \Sigma^-$ by testing and distinguishing the phenomenological models for the $\gamma n \to K^+ \Sigma^-$ reaction.

In 2019, the A2 Collaboration released the world first data on differential cross sections for the $\gamma n \to K^0 \Sigma^0$ reaction from threshold up to center-of-mass energy $W=1855$ MeV \cite{A2:2018doh}. Very recently, the BGOOD Collaboration reported the differential cross sections for this reaction from threshold up to $W=2400$ MeV \cite{BGOOD:2021oxp}. Since the hadronic and electromagnetic vertices are exactly the same in both the $\gamma n \to K^+ \Sigma^-$ and $\gamma n \to K^0 \Sigma^0$ reactions except for some possible isospin factors, the differential cross-section data from the A2 and BGOOD Collaborations for $\gamma n \to K^0 \Sigma^0$ can be, in principle, used to test the theoretical models for $\gamma n \to K^+ \Sigma^-$. Furthermore, it is expected that a combined analysis of the data for these two reactions would provide more experimental constraints on theoretical models to disentangle the resonance hadronic and electromagnetic couplings, and to achieve a better understanding of the reaction dynamics of the $K\Sigma$ photoproduction off neutron.

It is worthy to mention that except for the two most recent works of Refs.~\cite{Wei:2022nqp,Byd2021} which have been devoted to analyze all the available data \cite{Kohri:2006yx, AnefalosPereira:2009zw, CLAS-beam, Zachariou:2020kkb} for $K^+\Sigma^-$ photoproduction, in literature several works \cite{Vancraeyveld:2009qt,Mart:2019fau,Clymton:2021wof,Luthfiyah:2021yqe} have been devoted to analyze the LEPS data published in 2006 \cite{Kohri:2006yx} and CLAS data published in 2010 \cite{AnefalosPereira:2009zw} for $K^+\Sigma^-$ photoproduction and the A2 data published in 2019 \cite{A2:2018doh} for $K^0\Sigma^0$ photoproduction. In Ref.~\cite{Vancraeyveld:2009qt}, the LEPS data \cite{Kohri:2006yx} for $K^+\Sigma^-$ photoproduction were analyzed within a Regge-plus-resonance model. The data from Refs.~\cite{Kohri:2006yx,AnefalosPereira:2009zw} for $K^+\Sigma^-$ photoproduction and the data from Ref.~\cite{A2:2018doh} for $K^0\Sigma^0$ photoproduction together with the data for $K^+\Sigma^0$ and $K^0\Sigma^+$ photoproduction reactions were analyzed in isobar models in a series works of Refs.~\cite{Mart:2019fau,Clymton:2021wof,Luthfiyah:2021yqe}, where the contributions from resonances with spin from $1/2$ up to $15/2$ were considered.

In the present paper, we perform a combined analysis of the $\gamma n \to K^0 \Sigma^0$ and $\gamma n \to K^+ \Sigma^-$ reactions within the model constructed in our pervious work \cite{Wei:2022nqp} by further taking into account the differential cross-section data from the A2 \cite{A2:2018doh} and BGOOD \cite{BGOOD:2021oxp} Collaborations for $\gamma n \to K^0 \Sigma^0$. The purpose is to implement more experimental constraints on the theoretical model and thus to achieve a better understanding of the reaction dynamics of the $K\Sigma$ photoproduction off neutron, especially, to learn more about the resonances that couple significantly to these two reactions.

The paper is organized as follows. In Sec.~\ref{Sec:formalism}, we briefly review the framework of the theoretical model. The results are shown and discussed in Sec.~\ref{Sec:results}. The summary and conclusions are given in Sec.~\ref{sec:summary}.

\section{Formalism}  \label{Sec:formalism}

\begin{figure}[tbp]
\centering
{\vglue 0.15cm}
\subfigure[~$s$ channel]{
\includegraphics[width=0.45\columnwidth]{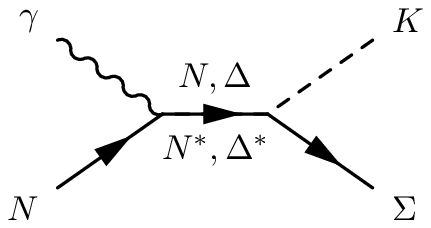}}  {\hglue 0.4cm}
\subfigure[~$t$ channel]{
\includegraphics[width=0.45\columnwidth]{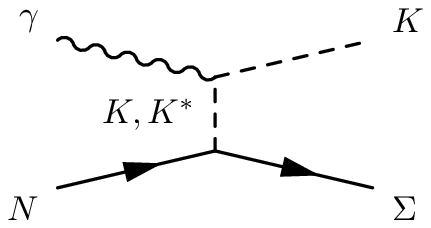}} \\[6pt]
\subfigure[~$u$ channel]{
\includegraphics[width=0.45\columnwidth]{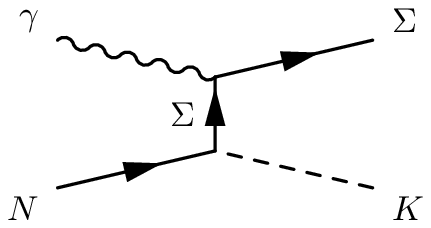}} {\hglue 0.4cm}
\subfigure[~Interaction current]{
\includegraphics[width=0.45\columnwidth]{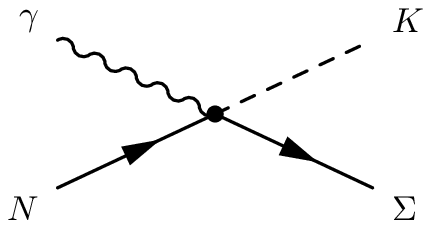}}
\caption{Generic structure of the amplitude for $\gamma N \to K \Sigma $. Time proceeds from left to right.}
\label{FIG:feymans}
\end{figure}

\begin{table*}[htb]
\caption{Fitted values of resonant parameters and the extracted resonance decay branching ratios obtained by fit A. The asterisks below the resonance names denote the overall rating of each resonance evaluated by RPP \cite{Zyla:2020zbs}. The values in the brackets blow resonances' masses, widths, and branching ratios are corresponding values advocated by RPP \cite{Zyla:2020zbs}. }
\begin{tabular*}{\textwidth}{@{\extracolsep\fill}ccccccc}
\hline\hline
 & $M_R$ [MeV] &$\Gamma_R$ [MeV] &$g^{(1)}_{RN\gamma} g_{R\Sigma K}$ &$g^{(2)}_{RN\gamma} g_{R\Sigma K}$ &$\beta_{N\gamma}$ [\%] &$\beta_{K\Sigma}$ [\%]  \\
\hline
$N(1710)1/2^+$   & $1692 \pm 1$  &$94 \pm 1$    &$1.475 \pm 0.010$    &&$\bm{0.01}$ &$0.018$      \\
$\ast\!\ast\!\ast\ast$  & $[1680\sim 1740]$   &$[80\sim 200]$ &&&$[0.0\sim 0.02]$   & [seen]     \\   \hline
$N(1880)1/2^+$  & $1860 \pm 1$  &$400 \pm 2$    &$1.605 \pm 0.001$    &&$\bm{0.05}$ &$21.363$       \\
$\ast\!\ast\!\ast$      & $[1830\sim 1930]$   &$[200\sim 400]$&&&$[0.002\sim 0.63]$ &$[10\sim 24]$    \\   \hline
$N(1895)1/2^-$  & $1920 \pm 1$  &$113 \pm 10$    &$0.015 \pm 0.001$    &&$\bm{0.005}$ &$14.085$       \\
$\ast\!\ast\!\ast\ast$  & $[1870\sim 1920]$   &$[80\sim 200]$ &&&$[0.003\sim 0.05]$ &$[6\sim 20]$     \\   \hline
$N(1900)3/2^+$  & $1931 \pm 1$  &$141 \pm 2$ &$-0.151 \pm 0.001$ &$0.083 \pm 0.005$ &$\bm{0.04}$ &$7.000$       \\
$\ast\!\ast\!\ast\ast$  & $[1890\sim 1950]$ &$[100\sim 320]$  &&&$[\textless 0.04]$ &$[3\sim 7]$     \\   \hline
$N(2060)5/2^-$  & $2051 \pm 2$  &$450 \pm 1$ &$-1.025 \pm 0.023$  &$-0.942 \pm 0.043$ &$\bm{0.037}$ &$1.948$       \\
$\ast\!\ast\!\ast$      & $[2030\sim 2200]$ &$[300\sim 450]$  &&&$[0.003 \sim 0.07]$ &$[1\sim 5]$     \\   \hline
$\Delta(1910)1/2^+$    & $1895 \pm 1$  &$200 \pm 2$  &$0.589 \pm 0.001$  && $\bm{0.02}$ &$14.000$       \\
$\ast\!\ast\!\ast\ast$  & $[1850\sim 1950]$ &$[200\sim 400]$  &&& $[0.0 \sim 0.02]$ &$[4\sim 14]$      \\   \hline
$\Delta(1920)3/2^+$ & $1870 \pm 1$ &$360 \pm 1$ &$-1.715 \pm 0.004$ &$0.365 \pm 0.066$ &$\bm{0.2}$ &$5.771$    \\
$\ast\!\ast\!\ast$      & $[1870\sim 1970]$ &$[240\sim 360]$  &&&   &$[2\sim 6]$     \\   \hline
\end{tabular*}
\label{tab:resonant parameters f1}
\end{table*}

\begin{table*}[htb]
\caption{Fitted values of cutoff parameters (in MeV) obtained by fit A.}
\begin{tabular*}{\textwidth}{@{\extracolsep\fill}cccccccccc}
\hline \hline
$\Lambda_{N(1710)}$ & $\Lambda_{N(1880)}$ & $\Lambda_{N(1895)}$ & $\Lambda_{N(1900)}$ & $\Lambda_{N(2060)}$ & $\Lambda_{\Delta(1910)}$ & $\Lambda_{\Delta(1920)}$ & $\Lambda_{K^\ast(K)}$  &  $\Lambda_{\Delta(N)}$  & $\Lambda_{\Sigma}$  \\ \hline
$1655 \pm 7$ & $1082 \pm 4$ &  $800 \pm 32$ & $2000 \pm 8$ & $1579 \pm 6$ & $1316 \pm 30$ & $1248 \pm 4$ &$589 \pm 1$ &$984 \pm 2$ &$1218 \pm 7$  \\ \hline
\hline
\end{tabular*}
\label{tab:cutoff_parameters f1}
\end{table*}

\begin{table*}[htb]
\caption{Fitted values of resonant parameters and the extracted resonance decay branching ratios obtained by fit B. The asterisks below the resonance names denote the overall rating of each resonance evaluated by RPP \cite{Zyla:2020zbs}. The values in the brackets blow resonances' masses, widths, and branching ratios are corresponding values advocated by RPP \cite{Zyla:2020zbs}. }
\begin{tabular*}{\textwidth}{@{\extracolsep\fill}ccccccc}
\hline\hline
 & $M_R$ [MeV] &$\Gamma_R$ [MeV] &$g^{(1)}_{RN\gamma} g_{R\Sigma K}$ &$g^{(2)}_{RN\gamma} g_{R\Sigma K}$ &$\beta_{N\gamma}$ [\%] &$\beta_{K\Sigma}$ [\%]  \\
\hline
$N(1710)1/2^+$   & $1692 \pm 1$  &$200 \pm 1$    &$2.677 \pm 0.056$    &&$\bm{0.01}$ &$0.020$      \\
$\ast\!\ast\!\ast\ast$  & $[1680\sim 1740]$   &$[80\sim 200]$ &&&$[0.0\sim 0.02]$   & [seen]     \\   \hline
$N(1880)1/2^+$  & $1830 \pm 1$  &$200 \pm 1$    &$0.306 \pm 0.001$    &&$\bm{0.005}$ &$21.275$       \\
$\ast\!\ast\!\ast$      & $[1830\sim 1930]$   &$[200\sim 400]$&&&$[0.002\sim 0.63]$ &$[10\sim 24]$    \\   \hline
$N(1895)1/2^-$  & $1920 \pm 1$  &$200 \pm 1$    &$-0.046 \pm 0.001$    &&$\bm{0.027}$ &$8.462$       \\
$\ast\!\ast\!\ast\ast$  & $[1870\sim 1920]$   &$[80\sim 200]$ &&&$[0.003\sim 0.05]$ &$[6\sim 20]$     \\   \hline
$N(1900)3/2^+$  & $1890 \pm 1$  &$235 \pm 3$ &$0.438 \pm 0.001$ &$-0.837 \pm 0.007$ &$\bm{0.04}$ &$9.668$       \\
$\ast\!\ast\!\ast\ast$  & $[1890\sim 1950]$ &$[100\sim 320]$  &&&$[\textless 0.04]$ &$[3\sim 7]$     \\   \hline
$N(2060)5/2^-$  & $2200 \pm 1$  &$450 \pm 2$ &$-0.000 \pm 0.001$  &$0.562 \pm 0.001$ &$\bm{0.01}$ &$1.994$       \\
$\ast\!\ast\!\ast$      & $[2030\sim 2200]$ &$[300\sim 450]$  &&&$[0.003 \sim 0.07]$ &$[1\sim 5]$     \\   \hline
$\Delta(1910)1/2^+$    & $1850 \pm 1$  &$200 \pm 1$  &$-0.854 \pm 0.012$  && $\bm{0.02}$ &$17.939$       \\
$\ast\!\ast\!\ast\ast$  & $[1850\sim 1950]$ &$[200\sim 400]$  &&& $[0.0 \sim 0.02]$ &$[4\sim 14]$      \\   \hline
$\Delta(1920)3/2^+$ & $1970 \pm 1$ &$286 \pm 16$ &$-0.318 \pm 0.006$ &$0.220 \pm 0.008$ &$\bm{0.04}$ &$3.262$    \\
$\ast\!\ast\!\ast$      & $[1870\sim 1970]$ &$[240\sim 360]$  &&&   &$[2\sim 6]$     \\   \hline
\end{tabular*}
\label{tab:resonant parametersf2}
\end{table*}

\begin{table*}[htb]
\caption{Fitted values of cutoff parameters (in MeV) obtained by fit B.}
\begin{tabular*}{\textwidth}{@{\extracolsep\fill}ccccccccccc}
\hline \hline
$\Lambda_{N(1710)}$ & $\Lambda_{N(1880)}$ & $\Lambda_{N(1895)}$ & $\Lambda_{N(1900)}$ & $\Lambda_{N(2060)}$ & $\Lambda_{\Delta(1910)}$ & $\Lambda_{\Delta(1920)}$ & $\Lambda_{K^\ast(K)}$  &  $\Lambda_{\Delta(N)}$  & $\Lambda_{\Sigma}$  \\ \hline
$876 \pm 4$ &$2066 \pm 3$ &$2039 \pm 46$ &$2100 \pm 2$ &$2000 \pm 4$ &$1436 \pm 25$ &$750 \pm 15$  &$567 \pm 1$ &$907 \pm 4$ &$1188 \pm 30$  \\ \hline
\hline
\end{tabular*}
\label{tab:cutoff_parametersf2}
\end{table*}

The tree-level effective Lagrangian model constructed for the $\gamma N \to K \Sigma$ reaction is diagrammatically depicted in Fig.~\ref{FIG:feymans}. One sees from this figure that the following reaction mechanisms are taken into account in constructing the $K\Sigma$ photoproduction amplitudes: Fig.~1(a) the $s$-channel $N$, $N^\ast$, $\Delta$, and $\Delta^\ast$ exchanges, Fig.~1(b) the $t$-channel $K$ and $K^\ast(892)$ exchanges, Fig.~1(c) the $u$-channel $\Sigma$ exchange, and Fig.~1(d) the interaction current which stands for other diagrams that do not have $s$-, $t$-, or $u$-channel poles and ensures the gauge invariance of the full photoproduction amplitudes. The gauge-invariant amputated photoproduction amplitudes of $\gamma N \to K \Sigma$ can be written as \cite{Haberzettl:1997,Haberzettl:2006bn,Huang:2012,Huang:2012xj}
\begin{equation}
M^{\mu} \equiv M^{\mu}_s + M^{\mu}_t + M^{\mu}_u + M^{\mu}_{\rm int},  \label{eq:amplitude}
\end{equation}
where the letter $\mu$ denotes the Lorentz index of the incoming photon $\gamma$. The terms $M^{\mu}_s$, $M^{\mu}_t$, $M^{\mu}_u$ and $M^{\mu}_{\rm int}$ stand for the amplitudes calculated from the $s$-channel mechanism, $t$-channel mechanism, $u$-channel mechanism and the interaction current, respectively. According to Refs.~\cite{Haberzettl:2006bn,Huang:2012xj}, the interaction current $M^{\mu}_{\rm int}$ can be modeled by a generalized contact current as
\begin{equation}
M^{\mu}_{\rm int} = \Gamma_{\Sigma NK}(q) C^\mu + M^{\mu}_{\rm KR} f_t,
\label{eq:Mint}
\end{equation}
with $q$ being the four-momentum of the outgoing $K$ meson and $f_t$ being the phenomenological form factor attaching to the amplitude of the $t$-channel $K$ exchange. Here, $\Gamma_{\Sigma NK}(q)$ is the vertex function of the $\Sigma NK$ interaction and reads
\begin{equation}
\Gamma_{\Sigma NK}(q) = g_{\Sigma NK}\gamma_5 \left(\lambda + \frac{1-\lambda}{2M_N} {\slashed q}\right),
\end{equation}
which is calculated from the following Lagrangian for the $\Sigma NK$ interaction,
\begin{equation}
{\cal L}_{\Sigma NK} = - g_{\Sigma NK}\bar{\Sigma}\gamma_{5} \left[\left(i\lambda + \frac{1-\lambda}{2M_N} \slashed{\partial}\right)K\right] N + \hc,  \label{eq:sigNK}
\end{equation}
with $\lambda$ being the mixing parameter for the pseudoscalar ($\lambda=1$) and pseudovector ($\lambda=0$) $\Sigma NK$ couplings.
The traditional Kroll-Ruderman term $M^{\mu}_{\rm KR}$ in Eq.~(\ref{eq:Mint}) reads
\begin{equation}
M^\mu_{KR} = -g_{\Sigma NK}\frac{1-\lambda}{2M_N} \gamma_5\gamma^\mu \tau Q_K,
\end{equation}
which is obtained from the following Lagrangian for the $\Sigma NK \gamma$ interaction,
\begin{eqnarray}
L_{\Sigma NK \gamma} &=& ig_{\Sigma NK}\frac{1-\lambda}{2M_N}\bar{\Sigma}\gamma_5\gamma^\mu A_\mu Q_K K\tau  N, \label{eq:sigNKgamma}
\end{eqnarray}
with $\tau$ being the isospin factor and $Q_K$ the electric charge of outgoing $K$ meson. The $C^\mu$ in Eq.~(\ref{eq:Mint}) is an auxiliary current introduced to ensure that the full photoproduction amplitude given in Eq.~(\ref{eq:amplitude}) satisfies the generalized Ward-Takahashi identity and thus is fully gauge invariant. According to Refs.~\cite{Haberzettl:1997,Haberzettl:2006bn,Huang:2012,Huang:2012xj}, for the $\gamma N \to K \Sigma$ reaction, the general prescription for $C^\mu$ is chosen as
\begin{eqnarray}
C^\mu &=&  - Q_{K} \tau \frac{f_t-\hat{F}}{t-q^2}  (2q-k)^\mu  - Q_{\Sigma} \tau \frac{f_{u}-\hat{F}}{u-p^{\prime 2}} (2p^{\prime}-k)^\mu  \nonumber   \\
      && - \tau Q_{N} \frac{f_{s}-\hat{F}}{s-p^{\prime 2}} (2p+k)^\mu,
\end{eqnarray}
with
\begin{equation}
\hat{F} = 1 - \hat{h} \left(1 -  \delta_t f_t\right) \left(1 -  \delta_u f_u\right) \left(1 -  \delta_s f_s\right),
\end{equation}
where $k$, $p$, and $p^\prime$ are the four-momenta for the incoming photon $\gamma$, incoming nucleon $N$, and outgoing $\Sigma$, respectively; $Q_{N}$ and $Q_{\Sigma}$  are the electric charges of incoming nucleon $N$ and outgoing $\Sigma$, respectively; $f_u$ and $f_s$ denote the phenomenological form factors attaching to the $u$-channel $\Sigma$ exchange and $s$-channel $N$ exchange, respectively; the constant $\delta_x=1$ for non-zero charges $Q_x$, and $\delta_x=0$ for zero charges $Q_x$; $\hat{h}=1$ is used for simplicity as usual \cite{Wang:2017tpe,Wang:2018vlv,Wei:2019}.

Note that the $t$-channel $K$ exchange, $u$-channel $\Sigma$ exchange as well as the interaction current contribute to the $\gamma n \to K^+ \Sigma^-$ reaction, but vanish in the $\gamma n \to K^0 \Sigma^0$ reaction due to the neutral charges of $K^0$ and $\Sigma^0$.

The amplitudes $M^{\mu}_s$, $M^{\mu}_t$, and $M^{\mu}_u$ can be straightforwardly obtained with the effective Lagrangians, resonance propagators, and phenomenological form factors that have been explicitly given in Sec.~II of Ref.~\cite{Wei:2022nqp}. We do not repeat these materials here for the sake of brevity.


\begin{figure}[htb]
\includegraphics[width=\columnwidth]{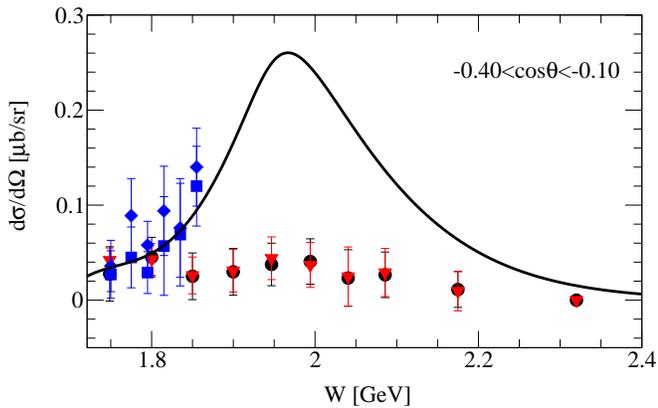}
\caption{Differential cross sections for $\gamma n \to K^0 \Sigma^0$ (black solid line) predicted from the model of Ref.~\cite{Wei:2022nqp}. The red inverted triangles and black circles denote the corresponding BGOOD data obtained by the fitting methods of RD and PS, respectively \cite{BGOOD:2021oxp}. The blue diamonds and squares denote the A2 data measured at $\cos\theta = -0.125$ and $\cos\theta = -0.375$, respectively \cite{A2:2018doh}. }
\label{fig:prediction}
\end{figure}

\begin{figure*}[htb]
\includegraphics[width=1.0\textwidth]{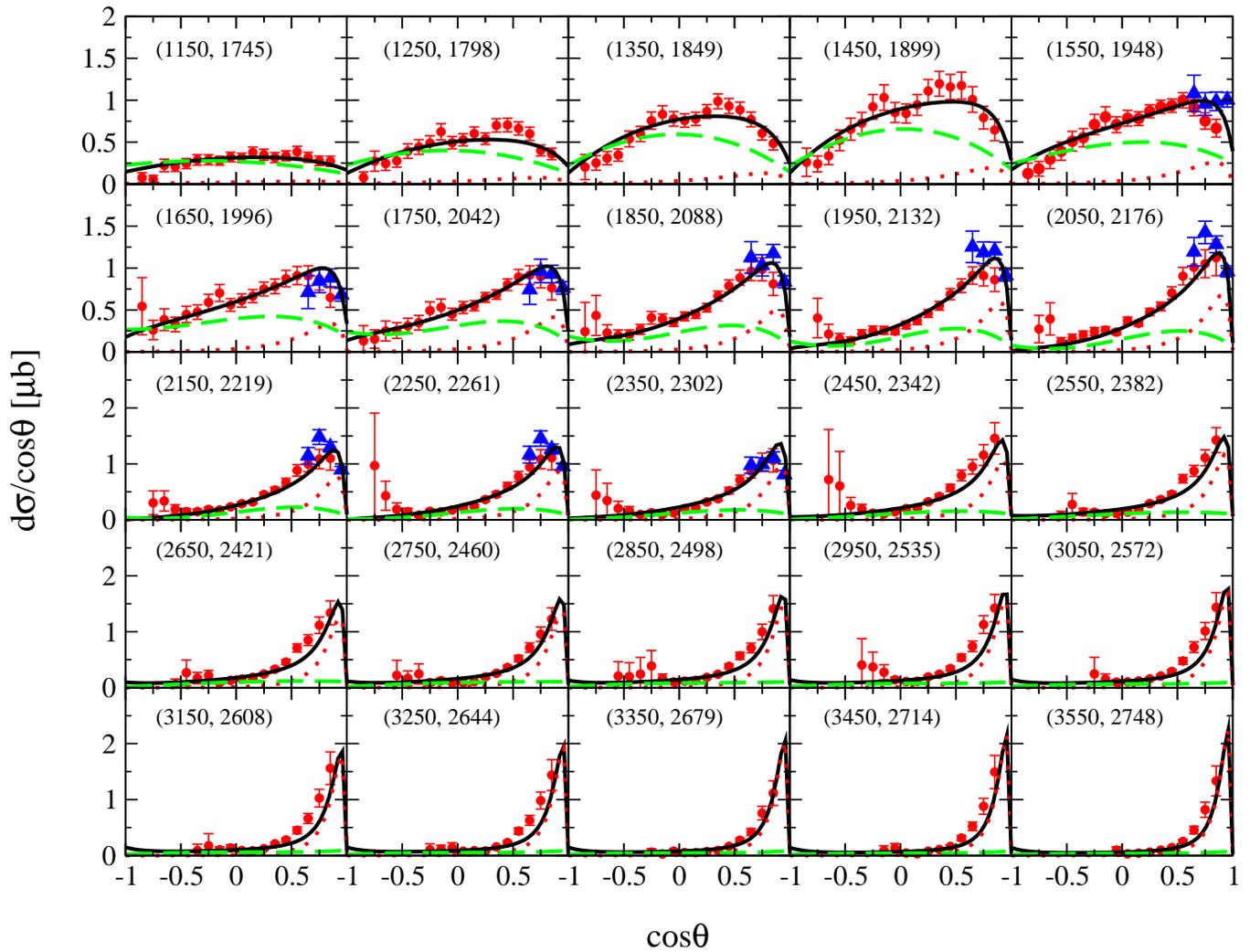}
\caption{Differential cross sections for $\gamma n \to K^+\Sigma^-$ obtained by fit A. The solid (black), dotted (red), and long-dashed (green) lines represent the results from the full calculation, the $t$-channel amplitudes ($K$ and $K^\ast(892)$ exchanges), and $s$-channel amplitudes ($N$, $N^\ast$, $\Delta$, and $\Delta^\ast$ exchanges), respectively. Data with red circles and blue triangles are taken from the CLAS Collaboration \cite{AnefalosPereira:2009zw} and the LEPS Collaboration \cite{Kohri:2006yx}, respectively. The numbers in parentheses denote the incident energies (left number) and the corresponding center-of-mass energies of the system (right number), in MeV. }
\label{fig:dsigf1}
\end{figure*}

\begin{figure*}[htb]
\includegraphics[width=1.0\textwidth]{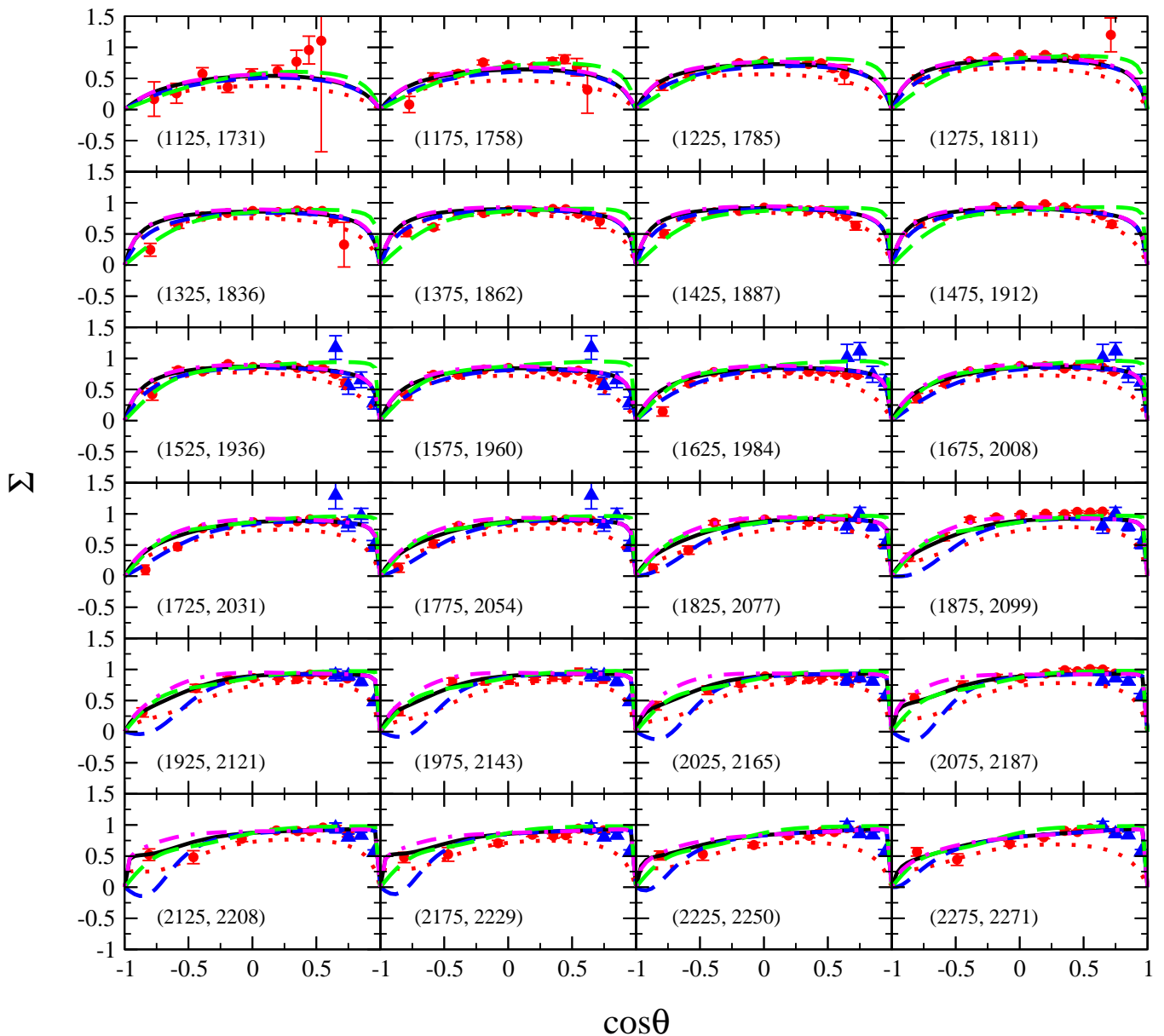}
\caption{Photo-beam asymmetries $\Sigma$ for $\gamma n \to K^+\Sigma^-$ obtained by fit A. The solid (black) lines represent the results from the full calculation. The dotted (red), dashed (blue), long-dashed (green), and dot-dashed (magenta) lines represent the results obtained by switching off the $t$-channel amplitudes ($K$ and $K^\ast(892)$ exchanges), $u$-channel amplitude ($\Sigma$ exchange), $s$-channel amplitudes ($N$, $N^\ast$, $\Delta$, and $\Delta^\ast$ exchanges), and interaction current, respectively. Data with red circles and blue triangles are taken from the CLAS Collaboration \cite{CLAS-beam} and the LEPS Collaboration \cite{Kohri:2006yx}, respectively. The numbers in parentheses denote the incident energies (left numbers) and the corresponding center-of-mass energies of the system (right numbers), in MeV.}
\label{fig:beamf1}
\end{figure*}

\begin{figure}[htb]
\includegraphics[width=\columnwidth]{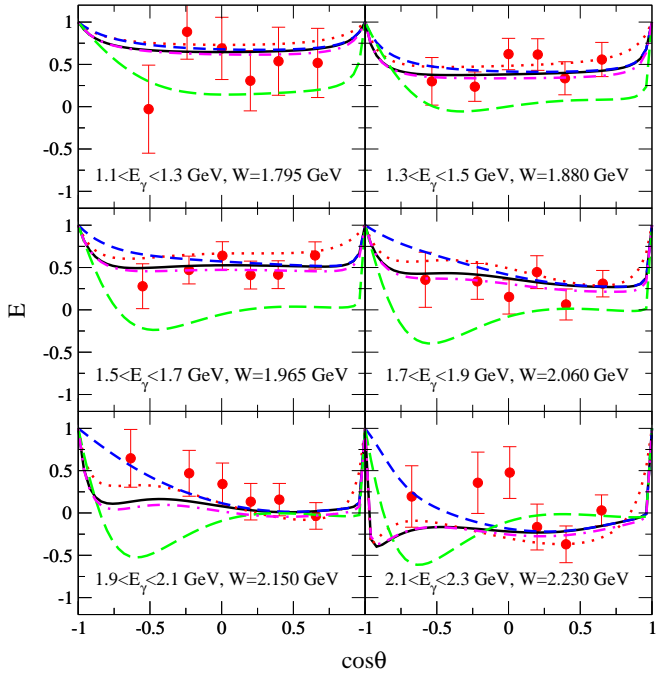}
\caption{Beam-target asymmetries $E$ for $\gamma n \to K^+\Sigma^-$ obtained by fit A. Notations for lines are the same as Fig.~\ref{fig:beamf1}. Data with red circles are taken from the CLAS Collaboration \cite{Zachariou:2020kkb}. $E_\gamma$ and $W$ denote the incident energies and the corresponding center-of-mass energies of the system, respectively.}
\label{fig:ef1}
\end{figure}

\begin{figure}[htb]
\includegraphics[width=\columnwidth]{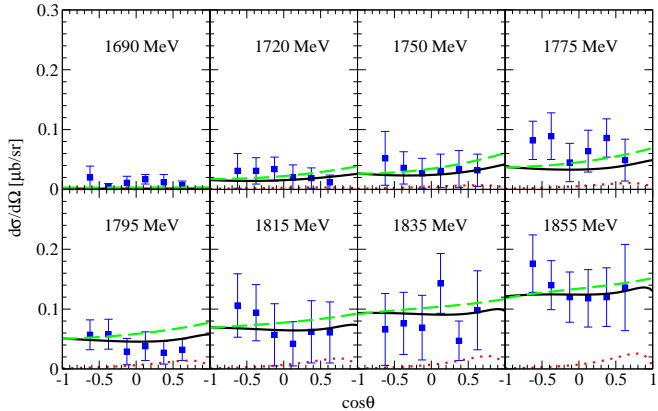}
\caption{Differential cross sections for $\gamma n \to K^0\Sigma^0$ obtained by fit A. The solid (black), dotted (red), and long-dashed (green) lines represent the results from the full calculation, the $t$-channel amplitude ($K^\ast(892)$ exchanges), and $s$-channel amplitudes ($N$, $N^\ast$, $\Delta$, and $\Delta^\ast$ exchanges), respectively. Data with blue squares are taken from the A2 Collaboration \cite{A2:2018doh}. The numbers in each subfigure denote center-of-mass energies (in MeV) of the system.}
\label{fig:mamif1}
\end{figure}

\begin{figure}[htb]
\includegraphics[width=\columnwidth]{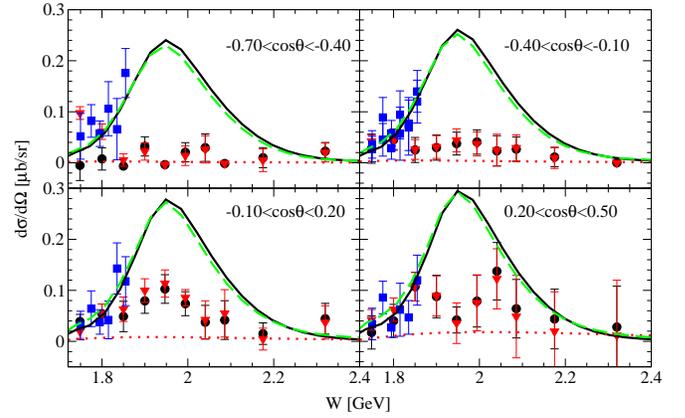}
\caption{Differential cross sections for $\gamma n \to K^0\Sigma^0$ obtained by fit A. The solid (black), dotted (red), and long-dashed (green) lines represent the results from the full calculation, the $t$-channel amplitude ($K^\ast(892)$ exchanges), and $s$-channel amplitudes ($N$, $N^\ast$, $\Delta$, and $\Delta^\ast$ exchanges), respectively. Data with red inverted triangles and black circles are taken from BGOOD Collaboration obtained by the fitting methods of RD and PS, respectively \cite{BGOOD:2021oxp}. Data with blue squares  are taken from the A2 Collaboration \cite{A2:2018doh}.}
\label{fig:bgoodf1}
\end{figure}

\begin{figure*}[htb]
\centering
\includegraphics[width=\columnwidth]{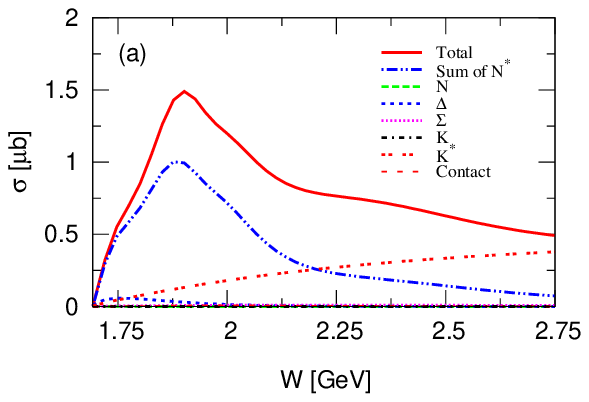}
\hspace{0.1 in}
\includegraphics[width=\columnwidth]{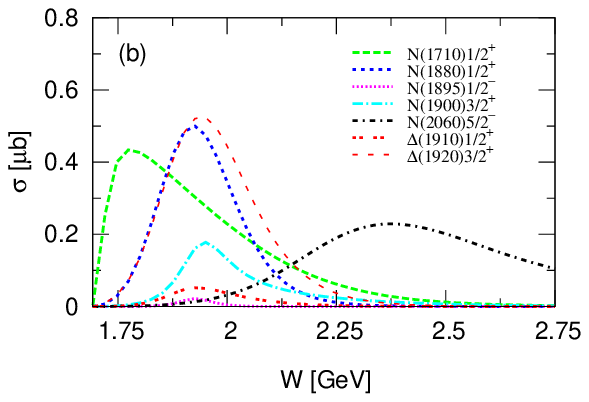}
\caption{Predicted total cross sections obtained by fit A for $\gamma n \to K^+\Sigma^-$ with contributions from individual terms.}
\label{fig:totalf1}
\end{figure*}

\begin{figure*}[htb]
\centering
\includegraphics[width=\columnwidth]{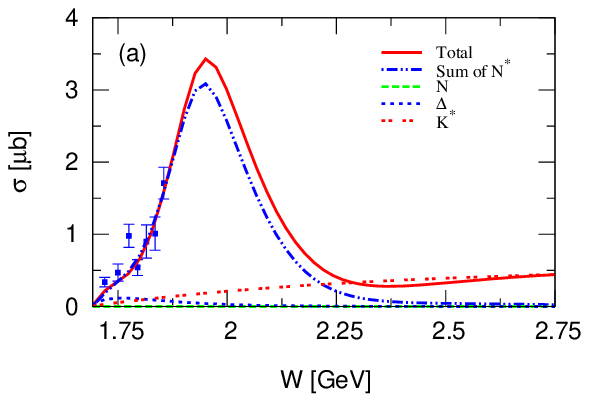}
\hspace{0.1 in}
\includegraphics[width=\columnwidth]{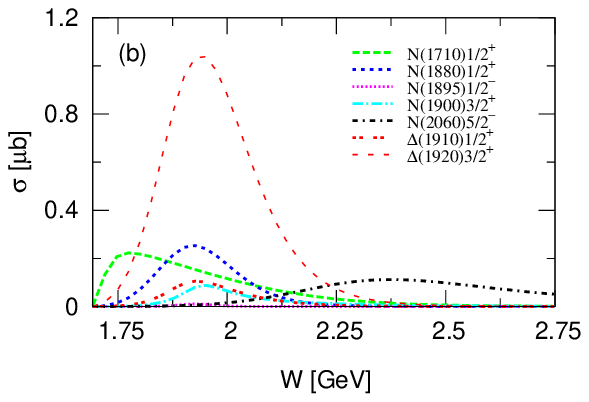}
\caption{Predicted total cross sections obtained by fit A for $\gamma n \to K^0\Sigma^0$ with contributions from individual terms. Data with blue squares are taken from the A2 Collaboration \cite{A2:2018doh}.}
\label{fig:totalc2f1}
\end{figure*}


\section{Results and discussion}   \label{Sec:results}


\begin{figure*}[htb]
\includegraphics[width=1.0\textwidth]{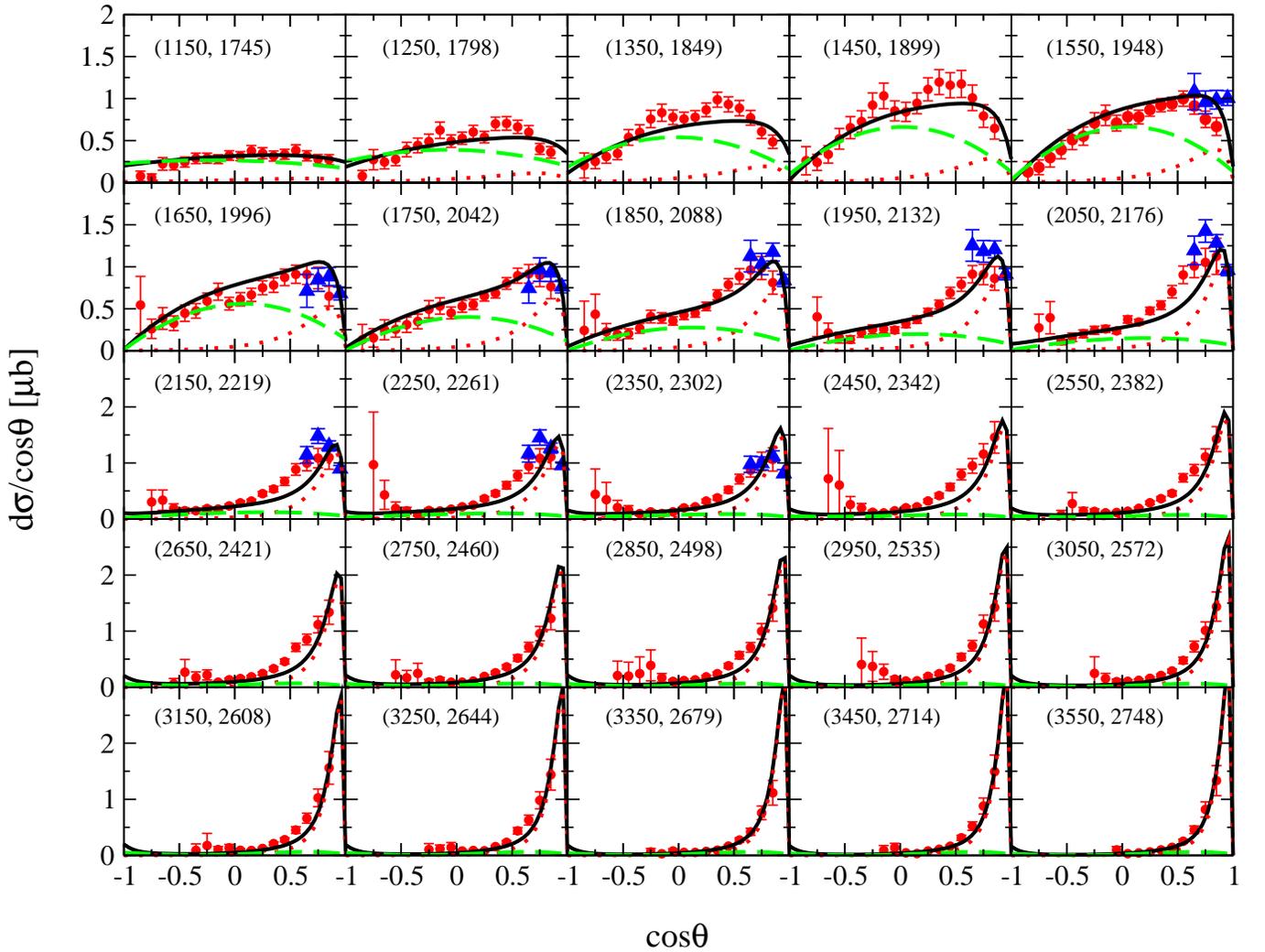}
\caption{Differential cross sections for $\gamma n \to K^+\Sigma^-$ obtained by fit B. Notations for lines and data are the same as Fig.~\ref{fig:dsigf1}.}
\label{fig:dsigf2}
\end{figure*}

\begin{figure*}[htb]
\includegraphics[width=1.0\textwidth]{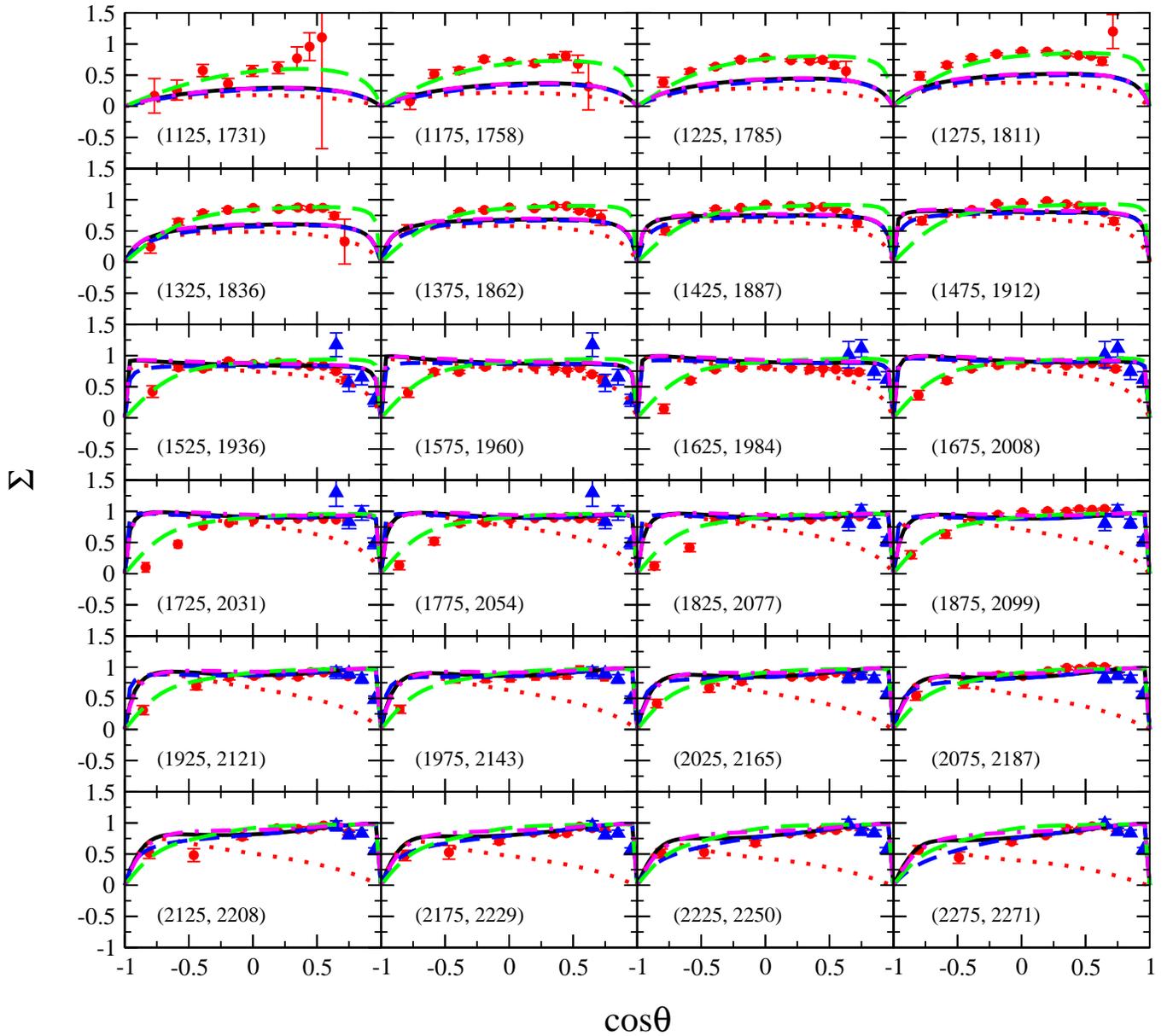}
\caption{Photo-beam asymmetries $\Sigma$ for $\gamma n \to K^+\Sigma^-$ obtained by fit B. Notations for lines and data are the same as Fig.~\ref{fig:beamf1}.}
\label{fig:beamf2}
\end{figure*}

\begin{figure}[htb]
\includegraphics[width=\columnwidth]{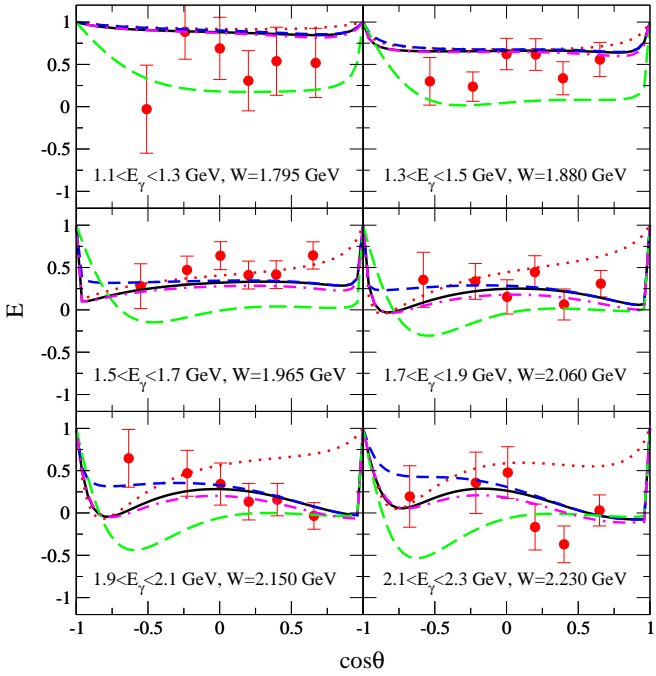}
\caption{Beam-target asymmetries $E$ for $\gamma n \to K^+\Sigma^-$ obtained by fit B. Notations for lines and data are the same as Fig.~\ref{fig:ef1}.}
\label{fig:ef2}
\end{figure}

\begin{figure}[htb]
\includegraphics[width=\columnwidth]{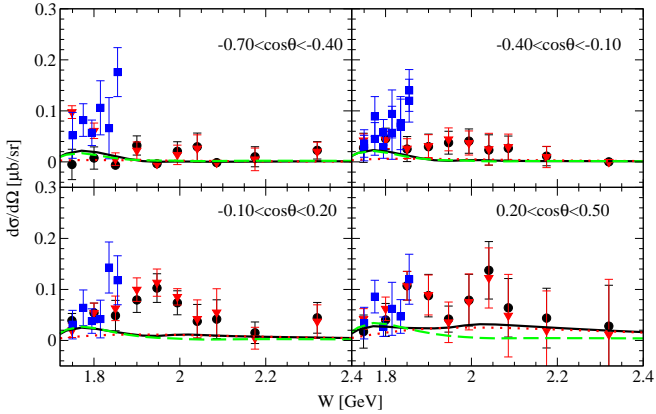}
\caption{Differential cross sections for $\gamma n \to K^0\Sigma^0$ obtained by fit B. Notations for lines and data are the same as Fig.~\ref{fig:bgoodf1}.}
\label{fig:bgoodf2}
\end{figure}

\begin{figure*}[htb]
\centering
\includegraphics[width=\columnwidth]{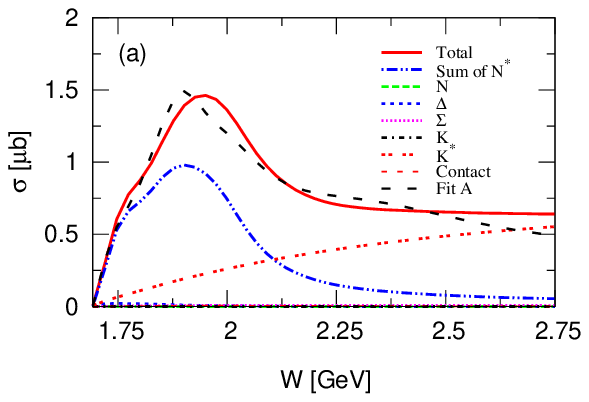}
\hspace{0.1 in}
\includegraphics[width=\columnwidth]{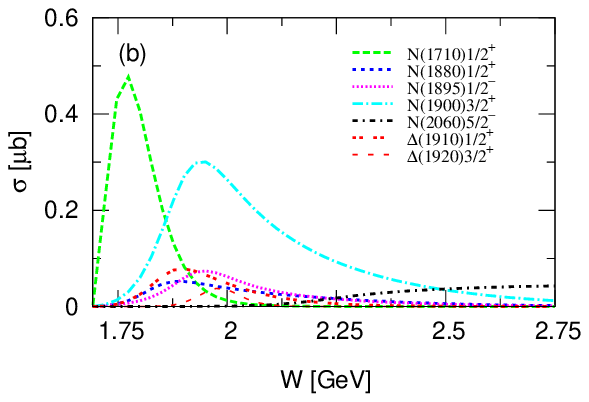}
\caption{Predicted total cross sections obtained by fit B for $\gamma n \to K^+\Sigma^-$ with contributions from individual terms. The black dashed line in the panel (a) represents the results from the full calculation of fit A.}
\label{fig:totalf2}
\end{figure*}

\begin{figure*}[htb]
\centering
\includegraphics[width=\columnwidth]{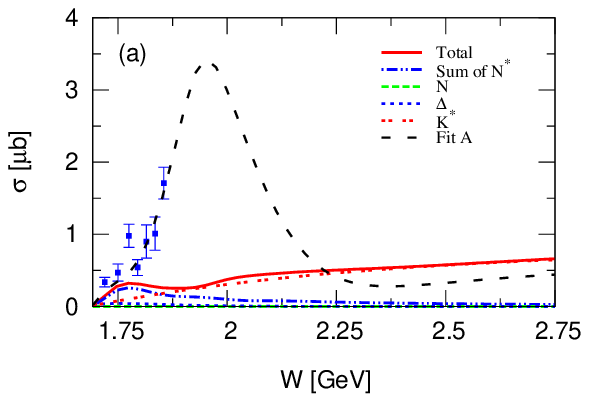}
\hspace{0.1 in}
\includegraphics[width=\columnwidth]{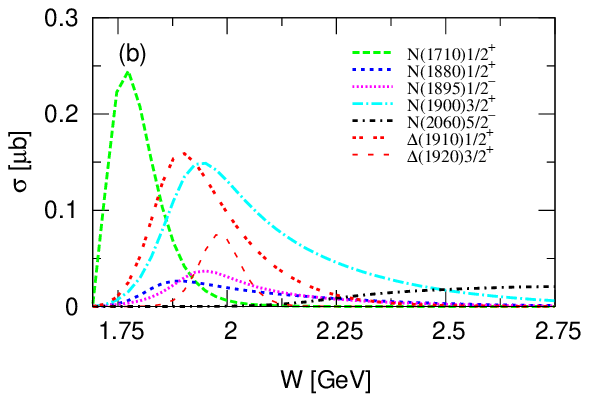}
\caption{Predicted total cross sections obtained by fit B for $\gamma n \to K^0\Sigma^0$ with contributions from individual terms. The black dashed line in the panel (a) represents the results from the full calculation of fit A. Data with blue squares are taken from the A2 Collaboration \cite{A2:2018doh}.}
\label{fig:totalc2f2}
\end{figure*}

\begin{figure}[htb]
\includegraphics[width=\columnwidth]{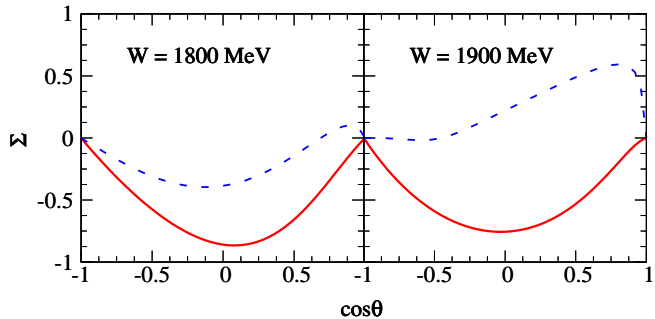}
\caption{Predictions of the photo-beam asymmetries $\Sigma$ obtained by Fit A (red solid lines) and Fit B (blue dashed lines) for $\gamma n \to K^0\Sigma^0$.}
\label{fig16}
\end{figure}

\begin{figure}[htb]
\includegraphics[width=\columnwidth]{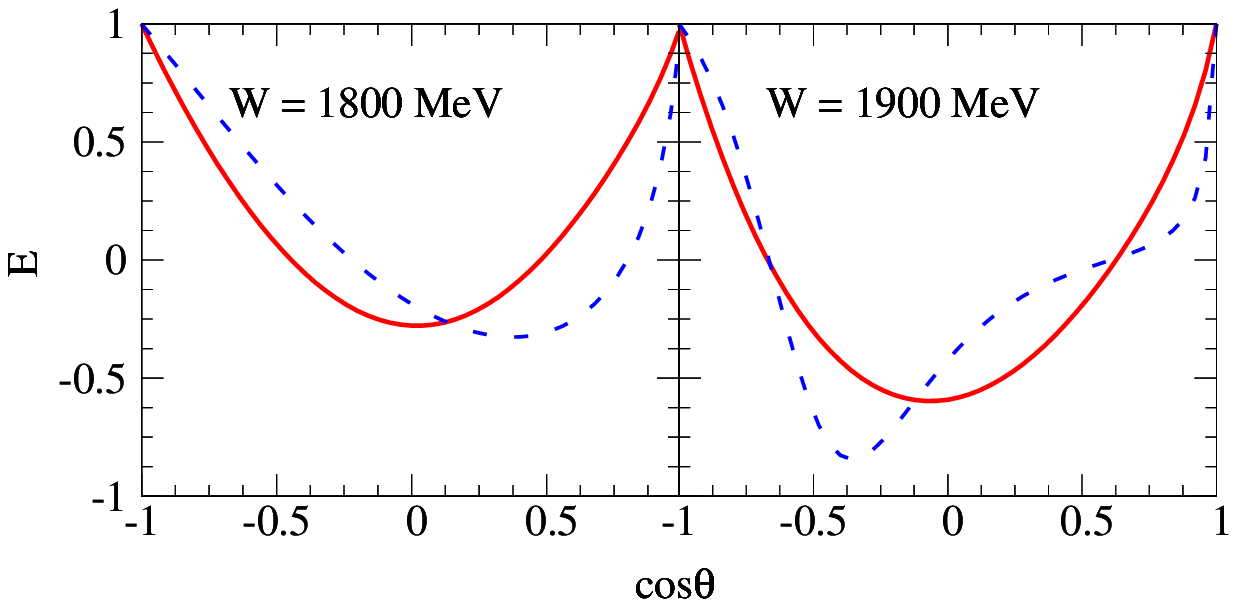}
\caption{Predictions of the beam-target asymmetries $E$ obtained by Fit A (red solid lines) and Fit B (blue dashed lines) for $\gamma n \to K^0\Sigma^0$.}
\label{fig17}
\end{figure}


The main purpose of the present work is to perform a combined analysis of the $\gamma n \to K^0 \Sigma^0$ and $\gamma n \to K^+ \Sigma^-$ reactions within the model constructed in our pervious work \cite{Wei:2022nqp} by further considering the differential cross-section data from the A2 and BGOOD Collaborations for the $\gamma n \to K^0 \Sigma^0$ reaction \cite{A2:2018doh,BGOOD:2021oxp}. No extra model parameter is needed to be introduced in constructing the amplitudes for the $\gamma n \to K^0 \Sigma^0$ reaction as the hadronic and electromagnetic vertices are exactly the same in both the $\gamma n \to K^+ \Sigma^-$ and $\gamma n \to K^0 \Sigma^0$ reactions except for some possible isospin factors. Therefore, the data for the $\gamma n \to K^0 \Sigma^0$ reaction can be used to test the theoretical model for the $\gamma n \to K^+ \Sigma^-$ reaction. Moreover, a combined analysis of the data for both the $\gamma n \to K^0 \Sigma^0$ and $\gamma n \to K^+ \Sigma^-$ reactions will implement more experimental constraints on the theoretical model to determine better the reaction amplitudes and to result in more reliable theoretical results.

Firstly we want to see how well the model of Ref.~\cite{Wei:2022nqp} can describe the data from the A2 and BGOOD Collaborations \cite{A2:2018doh,BGOOD:2021oxp} for $\gamma n \to K^0 \Sigma^0$. As an illustration, part of the model predictions of the differential cross sections for $\gamma n \to K^0 \Sigma^0$ is shown in Fig.~\ref{fig:prediction} with a comparison to the corresponding data from the A2 and BGOOD Collaborations \cite{A2:2018doh, BGOOD:2021oxp}. Here, the prediction of the model results (black solid line) is calculated at $\cos\theta = -0.25$. The red inverted triangles and black circles denote the corresponding BGOOD data at $-0.40 < \cos\theta < -0.10$ obtained by using the fitting methods of, respectively, RD and PS, which denote two different methods used to describe and subtract background \cite{BGOOD:2021oxp}. The blue diamonds and squares denote the A2 data measured at $\cos\theta = -0.125$ and $\cos\theta = -0.375$, respectively \cite{A2:2018doh}, which are located at the range of the angle measured by the BGOOD Collaboration. One can see from Fig.~\ref{fig:prediction} that the model results of Ref.~\cite{Wei:2022nqp} are able to qualitatively reproduce the data from the A2 Collaboration \cite{A2:2018doh}, but fail to describe the data from the BGOOD Collaboration \cite{BGOOD:2021oxp}. Furthermore, obvious discrepancies between the data sets are shown: the A2 data have an upward tendency as the center-of-mass energy increases with a maximum value $\approx 0.14$ ${\rm \mu b/sr}$, while the BGOOD data appear almost flat with values remaining blow $0.4$ ${\rm \mu b/sr}$.

Even though the BGOOD data are far from being satisfactorily described, the qualitative description of the A2 data is, to some extent, an advocate of the validity and feasibility of the theoretical model constructed in Ref.~\cite{Wei:2022nqp}. No matter what, a simultaneous analysis of the data for both the $\gamma n \to K^0 \Sigma^0$ and $\gamma n \to K^+ \Sigma^-$ reactions is anticipated to implement more experimental constraints on the theoretical model and thus to result in more reliable results. However, we have checked and found that, due to the discrepancies between the data sets of the A2 and BGOOD Collaborations, there is no way to obtain a unique fit if we consider both of them simultaneously. As we have no clear reason to discard one data set in favor of the other, we are therefore forced to consider them separately. In one case, we have considered the data from the A2 Collaboration \cite{A2:2018doh} for $\gamma n \to K^0 \Sigma^0$ and all the available data \cite{Kohri:2006yx, AnefalosPereira:2009zw, CLAS-beam, Zachariou:2020kkb} for $\gamma n \to K^+ \Sigma^-$, which is denoted as fit A. In the other case, we have considered the data from the BGOOD Collaboration \cite{BGOOD:2021oxp} for $\gamma n \to K^0 \Sigma^0$ and all the available data \cite{Kohri:2006yx, AnefalosPereira:2009zw, CLAS-beam, Zachariou:2020kkb} for $\gamma n \to K^+ \Sigma^-$, which is denoted as fit B.

In Ref.~\cite{Wei:2022nqp} contributions from the $N(1880)1/2^+$, $N(1895)1/2^-$, $N(1900)3/2^+$, $N(2060)5/2^-$, $\Delta(1910)1/2^+$, $\Delta(1920)3/2^+$, $\Delta(1950)7/2^+$, and $N(1710)1/2^+$ resonances have been taken into account to reproduce the available data for $\gamma n \to K^+ \Sigma^-$. Apart from the $N(1710)1/2^+$ which is marked as ``seen" in its decay branching ratio to the $K \Sigma$ channel, all the other considered resonances have sizable $K \Sigma$ branching ratios in RPP \cite{Zyla:2020zbs}. Nevertheless, the results of Ref.~\cite{Wei:2022nqp} have shown that the contributions from the high partial wave resonance $\Delta(1950)7/2^+$ are rather small. In the present combined analysis of $\gamma n \to K^0 \Sigma^0$ and $\gamma n \to K^+ \Sigma^-$, we have checked and found that the contributions from this resonance are even smaller. We thus do not consider this resonance in the present work.

The results obtained by fit A are shown in Fig.~\ref{fig:dsigf1}-\ref{fig:totalc2f1}. The resulted $\chi^2$ per degree of freedom, $\chi^2/N$, is $3.21$ with $N = 828$. The fitted values of resonant parameters as well as the extracted resonance decay branching ratios are listed in Table~\ref{tab:resonant parameters f1}. Here, the number of asterisks below each resonance names denote the overall status of this resonance evaluated by RPP \cite{Zyla:2020zbs}. The masses and widths of resonances have been varied within the ranges (values in the brackets below the resonance masses and widths) advocated by RPP \cite{Zyla:2020zbs} to fit the data for $\gamma n \to K^0\Sigma^0$ and $\gamma n \to K^+\Sigma^-$. In the present tree-level calculations, the calculated results are only sensitive to the products of the electromagnetic and hadronic coupling constants, and we, therefore, list the values of these products instead of individual electromagnetic and hadronic coupling constants. The last two columns of Table~\ref{tab:resonant parameters f1} show the extracted decay branching rations of $R \to N\gamma$ and $R \to K\Sigma$, respectively. Here, the values in  brackets denote the ranges of the corresponding branching ratios available in RPP \cite{Zyla:2020zbs}. For each resonance, its electromagnetic branching ratio is fixed at a selected value (in bold font) within the range advocated by RPP \cite{Zyla:2020zbs} and its hadronic branching ratio is then extracted from the corresponding products of the resonance electromagnetic and hadronic coupling constants. One can see from Table~\ref{tab:resonant parameters f1} that, for all the resonances, the electromagnetic branching ratios and the extracted hadronic branching ratios are all consistent to those advocated by RPP \cite{Zyla:2020zbs}. The corresponding fitted values of cutoff parameters are listed in Table~\ref{tab:cutoff_parameters f1}.

The differential cross sections for $\gamma n \to K^+\Sigma^-$ obtained by fit A (black-solid lines) are shown in Fig.~\ref{fig:dsigf1} and compared to the data from the CLAS \cite{AnefalosPereira:2009zw} Collaboration (red circles) and LEPS \cite{Kohri:2006yx} Collaboration (blue triangle). Individual contributions from various reaction dynamics are also shown. The red-dotted and green-long-dashed lines represent the results from the $t$-channel mechanisms ($K$ and $K^\ast(892)$ exchanges) and $s$-channel mechanisms ($N$, $N^\ast$, $\Delta$, and $\Delta^\ast$ exchanges), respectively. Note that, contributions from the interaction current and the $u$-channel mechanism are too small to be clearly seen with the scale used, and thus they are not plotted in this figure. The numbers in the parentheses denote the corresponding incident energies (left numbers) and the corresponding center-of-mass energies of the system (right numbers), in MeV. One can see that, the differential cross section data for $\gamma n \to K^+\Sigma^-$ can be well described with the model results. In the lower energy region, contributions from the $s$-channel mechanisms dominate the reaction, while contributions from the $t$-channel mechanisms become more and more prominent as the center-of-mass energy increases and finally dominate the reaction in the higher energy region. In particular, contributions from the $t$-channel $K^\ast(892)$ exchange account for the forward-angle-peaked behavior exhibited at higher energies.

The beam asymmetries $\Sigma$ (black-solid lines) and beam-target asymmetries $E$ (black-solid lines) for $\gamma n \to K^+\Sigma^-$ obtained by fit A are shown and compared to the data \cite{Kohri:2006yx,CLAS-beam,Zachariou:2020kkb} in Fig.~\ref{fig:beamf1} and Fig.~\ref{fig:ef1}, respectively. Here, the red-dotted, blue-dashed, green-long-dashed, and magenta-dot-dashed lines denote the results obtained by switching off the $t$-channel amplitudes ($K$ and $K^\ast(892)$ exchanges), $u$-channel amplitude ($\Sigma$ exchange), $s$-channel amplitudes ($N$, $N^\ast$, $\Delta$, and $\Delta^\ast$ exchanges), and the interaction current, respectively, in the full photoproduction amplitudes. Overall, the agreement of the model results with the data is reasonable. For the photo-beam asymmetries $\Sigma$ in Fig.~\ref{fig:beamf1}, contributions from the $s$- and $t$-channel mechanisms have significant effects. Contributions from the $u$-channel mechanism have noticeable effects at higher energies and backward angles, while the effects from the interaction current is quite small. For the beam-target asymmetries $E$ in Fig.~\ref{fig:ef1}, contributions from the $s$-channel mechanisms have dominant effects. The effects from the $t$-and the $u$-channel mechanisms are significant, while the effects from the interaction current are found to be negligible, just as the situation seen in the differential cross sections and photo-beam asymmetries $\Sigma$.

The results for the differential cross sections for $\gamma n \to K^0\Sigma^0$ obtained by fit A (black-solid lines) are shown in Fig.~\ref{fig:mamif1} and compared to the data from the A2 Collaboration \cite{A2:2018doh}. One can see that, the experimental data can be well described and the contributions from the $s$-channel mechanisms dominate this reaction.

In Fig.~\ref{fig:bgoodf1}, we also compare the differential cross sections for $\gamma n \to K^0\Sigma^0$ obtained by fit A (black-solid lines) to the data from the BGOOD Collaboration \cite{BGOOD:2021oxp}. The data from the A2 Collaboration \cite{A2:2018doh} measured at the same ranges of the meson polar angle are also shown for comparison. One can see that, the model results of fit A can well reproduce the data from the A2 Collaboration \cite{A2:2018doh} but fail to describe the data from the BGOOD Collaboration \cite{BGOOD:2021oxp}.

The predictions of the total cross sections for $\gamma n \to K^+\Sigma^-$ and $\gamma n \to K^0\Sigma^0$ obtained by fit A are shown in Fig.~\ref{fig:totalf1} and Fig.~\ref{fig:totalc2f1}, respectively. Clearly, one can see that in both reactions, the $s$-channel resonance exchanges dominate the reactions in the lower energy region, and the $t$-channel $K^\ast(892)$ exchange become more and more prominent as the center-of-mass energy increases and finally dominates the reactions in the higher energy region. In the lower energy region, the dominant $s$-channel contributions in both reactions are from the $N(1710)1/2^+$, $N(1880)1/2^+$, $N(1900)3/2^+$, and $\Delta(1920)3/2^+$ resonances. We remark that, even though the dominant resonances in $\gamma n \to K^+\Sigma^-$ found here are the same as those in Ref.~\cite{Wei:2022nqp}, the individual contributions from each resonance are noticeably different in these two works. In particular, the contributions from the $N(1710){1/2}^+$ resonance are much narrower and located at lower energy region in the present work due to the much smaller fit value of the cutoff parameter as listed in Table~\ref{tab:cutoff_parameters f1}. The contributions from both the $N(1880)1/2^+$ and $\Delta(1920)3/2^+$ resonances are much stronger in the present work mainly due to the larger branching ratios derived from the fitted resonance couplings as listed in Table~\ref{tab:resonant parameters f1}. The contributions from the $N(1900)3/2^+$ resonance in the present work are similar to those in Ref.~\cite{Wei:2022nqp}. The $N(2060){5/2}^-$ resonance contributes significantly at higher energy region in both works, while its contributions are a little bit more stronger in the present work due to the larger fit value of the cutoff parameter and bigger branching ratio derived from the fitted coupling constants. Nevertheless, the coherent sum of all resonance contributions in the $\gamma n \to K^+\Sigma^-$ reaction in fit A of the present work is similar to that in the fit of Ref.~\cite{Wei:2022nqp}. Note that the resulted $\chi^2/N$ for fit A in the present work is $3.37$, bigger than the value $3.11$ in Ref.~\cite{Wei:2022nqp}, indicating a little bit worse fitting quality of fit A in the present work than the fit in Ref.~\cite{Wei:2022nqp}, which is reasonable as the data from the A2 Collaboration \cite{A2:2018doh} implement additional constraints on the theoretical model.

We now come to the discussions of the results obtained by fit B. The resulted $\chi^2$ per degree of freedom is $\chi^2/N = 15.81$ with $N = 819$, which is significantly larger than that of fit A. Note that, for the BGOOD data, the two methods (RD and PS) used to describe and subtract back ground show a good agreement \cite{BGOOD:2021oxp}, and we have included the data obtained by RD in our fits. The results obtained by fit B are shown in Fig.~\ref{fig:dsigf2}-\ref{fig:totalc2f2}. The fitted values of resonant parameters as well as the extracted resonance decay branching ratios are listed in Table~\ref{tab:resonant parametersf2}. One sees from Table~\ref{tab:resonant parametersf2} that the fitted values of the model parameters have changed significantly compared with those shown in Table~\ref{tab:resonant parameters f1}. The extracted electromagnetic branching ratios and hadronic branching ratios are consistent to the ranges advocated by RPP \cite{Zyla:2020zbs}, except for those of the $N(1900)3/2^+$ and $\Delta(1910)1/2^+$ resonances, which are a little bigger than the corresponding values advocated by RPP \cite{Zyla:2020zbs}. The fitted values of cutoff parameters are listed in Table~\ref{tab:cutoff_parametersf2}.

The differential cross sections for $\gamma n \to K^+\Sigma^-$ obtained by fit B (black-solid lines) are shown in Fig.~\ref{fig:dsigf2} and compared to the data from the CLAS \cite{AnefalosPereira:2009zw} Collaboration (red circles) and LEPS \cite{Kohri:2006yx} Collaboration (blue triangle). One can see that the differential cross-section data can be also well described by the results of fit B. Similar to the situation of fit A, contributions from the $s$- and $t$-channel mechanisms dominate the reaction in the lower and higher energy regions, respectively.

The results for the beam asymmetries $\Sigma$ (black-solid lines) and beam-target asymmetries $E$ (black-solid lines) for $\gamma n \to K^+\Sigma^-$ obtained by fit B are shown and compared to the data \cite{Kohri:2006yx,CLAS-beam,Zachariou:2020kkb} in Fig.~\ref{fig:beamf2} and Fig.~\ref{fig:ef2}, respectively. For the photo-beam asymmetries $\Sigma$ in Fig.~\ref{fig:beamf2}, noticeable discrepancies between the data and model results are observed at lower energies, resulting in relative large $\chi^2/N$ as the measured error bars in the data are quite small. Contributions from the $s$-channel and $t$-channel mechanisms have significant effects on photo-beam asymmetries in the lower energy region and higher energy region, respectively, while the effects from the $u$-channel mechanism and the interaction current are found to be negligible. For the beam-target asymmetries $E$ in Fig.~\ref{fig:ef2}, the agreement of the model results with the data is reasonable. Contributions from the $s$-channel mechanisms have the most strong effects in the whole energy region. In the higher energy region, the effects from $t$-channel mechanisms are significant at forward angles, and noticeable effects from the $u$-channel mechanism are also shown. Again, the effects from the interaction current are found to be negligible.

The differential cross sections for $\gamma n \to K^0\Sigma^0$ obtained by fit B (black-solid lines) are shown in Fig.~\ref{fig:bgoodf2} and compared to the data from the BGOOD Collaboration \cite{BGOOD:2021oxp}. The data from the A2 Collaboration \cite{A2:2018doh} measured at the same ranges of the meson polar angle are also shown for comparison. One sees that, the experimental data can be only qualitatively described. The bump structure shown at $W \approx 1900$ MeV and $-0.10 < \cos\theta < 0.20$ can not be described by the model results.

The predictions of the total cross section for $\gamma n \to K^+\Sigma^-$ and $\gamma n \to K^0\Sigma^0$ obtained by fit B are shown in Fig.~\ref{fig:totalf2} and Fig.~\ref{fig:totalc2f2}, respectively. One can see that in both reactions, contributions from the $s$-channel mechanisms dominate the reactions in the lower energy region, while the contributions from the $t$-channel mechanisms dominate the reactions in the higher energy region. Although the $t$-channel interactions provide similar contributions in both $\gamma n \to K^+\Sigma^-$ and $\gamma n \to K^0\Sigma^0$, the contributions from the $s$-channel resonance diagrams in $\gamma n \to K^0\Sigma^0$ are much smaller than those in $\gamma n \to K^+\Sigma^-$. Compared with fit A where the contributions from $s$-channel diagrams are mainly coming from the $N(1710)1/2^+$, $N(1880)1/2^+$, $N(1900)3/2^+$, $\Delta(1920)3/2^+$, and $N(2060)5/2^-$ resonances, the dominant $s$-channel contributions in fit B are from the $N(1710)1/2^+$, $N(1900)3/2^+$, and $\Delta(1910)1/2^+$ resonances. For the $\gamma n \to K^+\Sigma^-$ reaction, the coherent sum of all resonance contributions from fit B is similar to that from fit A. On the contrary, for the $\gamma n \to K^0\Sigma^0$ reaction, the coherent sum of all resonance contributions from fit B is less than $1/6$ of that from fit A. One also sees that the predicted total cross sections for $\gamma n \to K^+\Sigma^-$ are similar in fits A and B, while for $\gamma n \to K^0\Sigma^0$ they are quite different.

We remark that the predictions of the differential cross sections and various polarization observables obtained by Kaon-Maid for $\gamma n \to K^+\Sigma^-$ and $\gamma n \to K^0\Sigma^0$ are available online \cite{Kaon-Maid}. Actually, the Kaon-Maid's predictions of the differential cross sections and beam asymmetries $\Sigma$ for $\gamma n \to K^+\Sigma^-$ have been published in Ref.~\cite{Clymton:2021wof}, and those of the beam-target asymmetries $E$ for $\gamma n \to K^+\Sigma^-$ have been published in Ref.~\cite{Zachariou:2020kkb}. One notes that the predictions of Kaon-Maid for the differential cross sections of $\gamma n \to K^0\Sigma^0$ agree well with the BGOOD data. Nevertheless, as shown in Refs.~\cite{Zachariou:2020kkb,Clymton:2021wof}, noticeable discrepancies can be seen between the data and the predictions of Kaon-Maid for the results of the differential cross sections, beam asymmetries $\Sigma$, and beam-target asymmetries $E$ for $\gamma n \to K^+\Sigma^-$.

In brief, by considering different data sets from A2 and BGOOD Collaborations \cite{A2:2018doh,BGOOD:2021oxp} for $\gamma n \to K^0\Sigma^0$, the model results of fit A and fit B have led to different conclusions about the reaction mechanisms. In fit A where the data from A2 Collaboration are considered, all data can be well described by the theoretical model, while in fit B where the data from BGOOD Collaboration are considered, noticeable discrepancies between the data and model results are seen. At present, we have no clear reason to discard on data set in favor of the other. Thus, further precise measurements of data for $\gamma n \to K^0\Sigma^0$ are called on to disentangle the discrepancies between the data sets from the A2 and BGOOD Collaborations \cite{A2:2018doh,BGOOD:2021oxp}, and to further determine the reaction mechanisms for $K\Sigma$ photoproduction reactions. Since the polarization observables are more sensitive to the details of the reaction dynamics in general and thus are preferred by experimentalist, in Figs.~\ref{fig16} and \ref{fig17} the predictions of the photo-beam asymmetries $\Sigma$ and beam-target asymmetries $E$ obtained by Fit A and Fit B for the $\gamma n \to K^0\Sigma^0$ reaction at $W = 1800$ and $W = 1900$ MeV are given, which can be examined by the future experiments.

\section{Summary and conclusion}  \label{sec:summary}

The available data for the $\gamma n \to K^+\Sigma^-$ reaction have been analyzed in our previous work \cite{Wei:2022nqp} and in Ref.~\cite{Byd2021}. However, these two independent analyses have extracted different resonance contents and different reaction mechanisms for this reaction even they described the same sets of data.

Recently, the A2 \cite{A2:2018doh} and BGOOD \cite{BGOOD:2021oxp} Collaborations released the data on differential cross sections for the $\gamma n \to K^0 \Sigma^0$ reaction. Since the hadronic and electromagnetic vertices are exactly the same in the $\gamma n \to K^+ \Sigma^-$ and $\gamma n \to K^0 \Sigma^0$ reactions except for some possible isospin factors, the data for $\gamma n \to K^0 \Sigma^0$ can be used to test the theoretical models of the $\gamma n \to K^+ \Sigma^-$ reaction. A combined analysis of the data for the $\gamma n \to K^+ \Sigma^-$ and $\gamma n \to K^0 \Sigma^0$ reactions would provide more experimental constraints on the theoretical models and, therefore, allow us perform a better analysis of the reaction mechanism and extract more reliably the resonance contents and parameters.

In the present paper, we have examined the theoretical model constructed in our previous work \cite{Wei:2022nqp} for $\gamma n \to K^+\Sigma^-$ by comparing the predicted differential cross sections for $\gamma n \to K^0 \Sigma^0$ with the corresponding data, and it was found that only the A2 data can be qualitatively reproduced. We have then performed a combined analysis of the data for both the $\gamma n \to K^0 \Sigma^0$ and $\gamma n \to K^+ \Sigma^-$ reactions \cite{Kohri:2006yx, AnefalosPereira:2009zw, CLAS-beam, Zachariou:2020kkb, A2:2018doh, BGOOD:2021oxp}. Due to the inconsistencies of data, we have included the data from the A2 and BGOOD Collaborations separately in the fits.

In the case of including the A2 data, all the data for the $\gamma n \to K^0 \Sigma^0$ and $\gamma n \to K^+ \Sigma^-$ reactions can be well described. Contributions from the $s$- and $t$-channel mechanisms were found to dominate the reactions in the lower energy region and higher energy region, respectively. The dominant $s$-channel contributions in the lower energy region were found to be from the $N(1710)1/2^+$, $N(1880)1/2^+$, $N(1900)3/2^+$, and $\Delta(1920)3/2^+$ resonances.

In the case of including the BGOOD data, the data for the $\gamma n \to K^+ \Sigma^-$ reaction can be also reproduced, with the exception of some noticeable discrepancies in the descriptions of the beam asymmetries. The data on the differential cross sections from the BGOOD \cite{BGOOD:2021oxp} Collaboration for the $\gamma n \to K^0 \Sigma^0$ reaction can be only qualitatively described. The $s$- and $t$-channel mechanisms were also found to dominate the reactions in the lower energy region and higher energy region, respectively. The dominant $s$-channel contributions in the lower energy region were found to be from the $N(1710)1/2^+$, $N(1900)3/2^+$, and $\Delta(1910)1/2^+$ resonances. Further precise measurements of data for $\gamma n \to K^0\Sigma^0$ were called on to disentangle the discrepancies between the data sets from the A2 and BGOOD Collaborations.

\begin{acknowledgments}
The author N. C. Wei thanks K. Kohl, who kindly provided him with the experimental data for the present work. This work is partially supported by the National Natural Science Foundation of China under Grants No.~12175240, No.~12147153, and No.~11635009, the Fundamental Research Funds for the Central Universities, and the China Postdoctoral Science Foundation under Grants No.~2021M693141 and No.~2021M693142.
\end{acknowledgments}

\end{document}